\documentclass[%
preprint, showpacs,preprintnumbers,
 amsmath,amssymb,
 aps,
]{revtex4-1}
\usepackage{amsmath}
\usepackage{cases}
\usepackage{amssymb}
\usepackage{CJK}% Chinese character support
\usepackage{graphicx}% Include figure files
\usepackage{dcolumn}% Align table columns on decimal point
\usepackage{bm}% bold math

\begin{document}
\begin{CJK*}{GBK}{} % Use default fonts from CJK (see below)

\preprint{APS/123-QED}

\title{Isobaric yield ratios in heavy-ion reaction, and symmetry energy of neutron-rich nuclei at intermediate energy}% Force line breaks with \\

\author{Chun-Wang MA}\email{machunwang@126.com}\author{Fang WANG}
\affiliation{%
 Department of Physics, Henan Normal University, Xinxiang, 453007 China %\textbackslash\textbackslash
}%
\author{Yu-Gang MA}\email{ygma@sinap.ac.cn}
\affiliation{%
Shanghai Institute of Applied Physics, Chinese Academy of Sciences, Shanghai, 201800 China %\textbackslash\textbackslash
}%
\author{Chan JIN}\email{jinchan2010@yahoo.cn}
\affiliation{Institute of Biophysics, The Second Military Medical University, Shanghai, 200433 China%
% Institute of Biophysics, The Second Military Medical University, Shanghai, 200433 China %\textbackslash\textbackslash
}%

%\affiliation{
%\affiliation{

\date{\today}% It is always \today, today,

\begin{abstract}
The isobaric yield ratios of the fragments produced in the
neutron-rich $^{48}$Ca and $^{64}$Ni projectile fragmentation are
analyzed in the framework of a modified Fisher model. The
correlations between the isobaric yield ratios ($R$) and the energy coefficients in the Weisz\"{a}cker-Beth semiclassical mass
formula (the symmetry-energy term $a_{sym}$, the Coulomb-energy term $a_c$, and
the pairing-energy term $a_p$) and the difference between the chemical potential of neutron and proton ($\mu_n-\mu_p$) are investigated. Simple correlations between ($\mu_n-\mu_p$)/T, $a_c$/T, $a_{sym}$/T, and
$a_p$/T (where T is the temperature), and ln$R$ are obtained. It is suggested that
($\mu_n-\mu_p$)/T, $a_c$/T, $a_{sym}$/T, and $a_p$/T of neutron-rich
nuclei can be extracted using isobaric yield ratios for heavy-ion collisions at
intermediate energies.
\end{abstract}

\maketitle
\end{CJK*}

%\tableofcontents

%\section{introduction}
\section{Introduction}
\label{intro} The construction of a new generation of radioactive nuclear beam facilities has stimulated much research into isospin physics \cite{BALi08}. In heavy-ion
reactions at intermediate energy, multifragmentation of the
reaction system is generally observed in violent collisions and
there is evidence that both subsaturated and supersaturated
densities can be explored in such collisions
\cite{Kowa07,BALi03,YGMa99}. Work in this area has concentrated
on exploring the nuclear equation of state (EOS) and the
liquid-gas phase transition in nuclear matter \cite{Bona,GoodFisher,DasPhRep05,Gupta01}. Isotopic yields in heavy-ion collisions provide a good probe for studying the nature of the disassembling nuclear systems. Many
studies on fragment emission have attempted to use
fragment yield distributions, either singly or by comparison to
those of similar reactions, to explore the symmetry energy of the
emitting source at different densities and temperatures
\cite{Gupta01,Kowa07,HSXu,Tsang01,DasPhRep05,Botv02,Ono03,YGMa05,YGMa97,YGMa99,Fang07JPG,MaCW08,Fang10,SunXY10,MaCW09,MaCW08ijmpe}. The nuclear symmetry energy of a finite nucleus is an important parameter in the EOS of an asymmetric nucleus and in various process in astrophysics and nuclear astrophysics. But the symmetry energy is difficult to measure experimentally and there are large differences in the theoretical results between different models and even within the same model with different parameters~\cite{BALi08}.

In a recent work analyzing isobaric yields
\cite{Huang10}, the ratio of the symmetry-energy coefficient to
temperature, $a_{sym}/T$, as a function of fragment mass $A$
was studied in a modified Fisher model (MFM)
\cite{ModelFisher1,ModelFisher2}. The Coulomb-energy coefficient
to temperature ($a_c/T$), and the pairing-energy to
temperature ($a_p/T$) were also extracted at the same time. For the
symmetry-energy term, the extracted values from experiments
are in good agreement with those calculated for the final
fragments in the ground states. The pairing effect is clearly
observed in experiments and strongly supports the hypothesis that
the observed effect originates at the end of the statistical
cooling-down process of the excited fragments. A comparison between
the Coulomb coefficients extracted experimentally and those
calculated shows significant differences.

In this article, on the basis of the theory of a modified Fisher
model \cite{ModelFisher1,ModelFisher2}, which was adopted in
Ref.~\cite{Huang10}, the correlation between the logarithm of the
isobaric yield ratio $\mbox{ln}R$ and  $a_{sym}/T$, $a_c/T$,
$a_p/T$ and $(\mu_n-\mu_p)/T$ for fragments produced in 140 $A$ MeV
$^{48}$Ca + $^{9}$Be and $^{64}$Ni + $^{9}$Be reactions
(experimental data are taken from Refs.~\cite{Mocko1,Mocko2}) are
analyzed. Coefficients of the volume energy, the surface energy,
the Coulomb energy, the symmetry energy and the pairing energy in the Weisz\"{a}cker-Beth semiclassical mass
formula and $(\mu_n-\mu_p)/T$ for fragments will be extracted using these
correlations.

\section{Isobaric Yield Ratios in the Modified Fisher Model}

Following the modified Fisher model theory
\cite{DasPhRep05,ModelFisher1,ModelFisher2}, the yield of fragment with
mass number $A$ and $I = N - Z$, $Y(A,I)$ is given by
\begin{equation}
Y(A,I) = CA^{-\tau}exp\{[W(A,I)+\mu_{n}N+\mu_{p}Z]/T+Nln(N/A)+Zln(Z/A)\},
\end{equation}
where $C$ is a constant. The A$^{-\tau}$ term originates from the
entropy of the fragment, and the last two terms are from the
entropy contributions for the mixing of two substances in the
Fisher droplet model \cite{Fisher}. $\mu_n$ and $\mu_p$ are the
neutron and proton chemical potentials, respectively, and $W(A,I)$
is the free energy of the cluster at temperature $T$. $W(A,I)$ is
given by the generalized Weisz\"{a}cker-Beth semiclassical mass
formula \cite{Weiz,Bethe} at a given temperature $T$ and density
$\rho$:
\begin{equation}\label{W}
W(A,I) = -E_{sym}-a_c(\rho,T)Z(Z-1)/A^{1/3} \\
+a_v(\rho,T)A-a_s(\rho,T)A^{2/3}-\delta(N-Z),
\end{equation}
where the indices $v,s,c,$ and $sym$ represent volume, surface,
Coulomb, and symmetry energies, respectively. The symmetry energy $E_{sym}$, can be divided into a volume-symmetry term $S_v$ and a surface-symmetry term $S_s$, i.e. $E_{sym}=S_v(\rho, T)+S_s(\rho,T)$, in which $S_v(\rho,T)=a_{vsym}(\rho, T)I^2/A$ and $S_s(\rho,T)=a_{ssym}(\rho,T)I^2A^{2/3}$ \cite{SJLee10,Nikolov11}. Following the
semiempirical mass formulas, the pairing energy $\delta(N,Z)$ is
given by~\cite{Green}
\begin{equation} \label{paring}
\delta(N,Z) = \left\{ \begin{aligned}
         &a_p(\rho,T)/A^{1/2} &\mbox{(odd-odd)}, \\
         &0 &\mbox{(even-odd)},\\
         &-a_p(\rho,T)/A^{1/2} &\mbox{(even-even)},
                          \end{aligned} \right.
                          \end{equation}

The yield ratio for fragments, $R(I + 2, I, A)$, between isobars differing by 2 units in $I$ is defined as
\begin{multline}\label{ratiodef}
R(I+2,I,A) = Y(A,I+2)/Y(A,I)\\
=\mbox{exp}\{[W(I+2,A)-W(I,A)+(\mu_n-\mu_p)]/T+S_{mix}(I+2,A)-S_{mix}(I,A)\},
\end{multline}
where $S_{mix}(I,A)=Nln(N/A)+Zln(Z/A)$. To simplify the
description, the density and temperature dependence of the
coefficients in Eq. (\ref{W}) is written as $a_i = a_i(\rho,T)$
(where $i = v, s, c, ssym, vsym$ and $p$ represent the volume energy, the surface energy, the Coulomb energy, the surface-symmetry energy, the volume symmetry energy, and the paring energy, respectively). The temperature dependence of $a_i$ at low T has been studied \cite{SJLee10}.

Inserting Eq.~(\ref{W}) into Eq.~(\ref{ratiodef}), one gets
\begin{eqnarray}\label{rdif2}
R(I+2,I,A)=&\mbox{exp}\{[(\mu_n-\mu_p)-4a_{sym}(I+1)/A+2a_c(Z-1)/A^{1/3}     \nonumber\\
           &-\delta(N+1,Z-1)+\delta(N,Z)]/T+\Delta(I+2,I,A)\},
\end{eqnarray}
where $\Delta(I+2,I,A) = S_{mix}(I+2,A)-S_{mix}(I,A)$. Similarly,
one can define the fragment yield ratio $R(I + 4, I, A)$ between
isobars differing by 4 units in $I$ following Eq.~(\ref{ratiodef})
as
\begin{multline}\label{ratiodef4}
R(I+4,I,A)=Y(A,I+4)/Y(A,I)\\
=\mbox{exp}\{[W(I+4,A)-W(I,A)+2(\mu_n-\mu_p)]/T+S_{mix}(I+4,A)-S_{mix}(I,A)\},
\end{multline}
and inserting Eq.~(\ref{W}) into Eq.~(\ref{ratiodef4}), one
gets
\begin{eqnarray}\label{rdif4}
R(I+4,I,A)=&\mbox{exp}\{[2(\mu_n-\mu_p)-8a_{sym}(I+2)/A+2a_c(2Z-3)/A^{1/3}     \nonumber\\
           &-\delta(N+2,Z-2)+\delta(N,Z)]/T+\Delta(I+4,I,A)\}.
\end{eqnarray}
Equations~(\ref{ratiodef4}) and (\ref{rdif4}) assume that $a_s$ and $a_v$ are the same for isobars, and omit the surface-symmetry-energy term as Ref. \cite{Huang10}. In this case, $a_{vsym}$ is written as $a_{sym}$ according to Ref. \cite{Huang10}.

For isobars with $I = -1$ and $I = 1$ (which are mirror nuclei),
$\Delta(1,-1,A) = 0$, and the contributions from the symmetry term
and the mixing entropy term in Eq.~(\ref{rdif2}) drop out and, the
pairing term also cancels out because these isobars are even-odd
nuclei. Taking the logarithm of the resultant equation, one
obtains
\begin{equation}\label{Rmirror1}
\mbox{ln}[R(1,-1,A)]=[(\mu_n-\mu_p)+2a_c(Z-1)/A^{1/3}]/T.
\end{equation}

Following Eq.~(\ref{ratiodef}), an isobar with odd \textit{"I"} is an
odd-even nucleus and the pairing energy is zero, so one gets
\begin{equation}\label{sym1}
\mbox{ln}[R(I+2,I,A)]=[(\mu_n-\mu_p)-8a_{sym}/A+2a_c(Z-1)/A^{1/3}]/T+\Delta(I+2,I,A).\\
\end{equation}
Considering ratios of  isobars with \textit{"I-2"}, \textit{"I"},
and \textit{"I+2"}, and assuming isobars with \textit{"I"} in
$R(I,I-2,A)$ and isobars with \textit{I+2} in $R(I+2,I,A)$ are
isotopes, one can reach
\begin{equation}\label{sym2}
(8a_{sym}/A+2a_c/A^{1/3})/T=\mbox{ln}[R(I,I-2,A)]-\mbox{ln}[R(I+2,I,A)]-\Delta(I,I-2,A)+\Delta(I+2,I,A). \\
\end{equation}
In Eq.~(\ref{sym2}), isobars with \textit{"I-2"} are taken as the
reference nuclei.

Following Eq.~(\ref{ratiodef4}) and taking isobars with
\textit{"I"} as the reference nuclei, one gets
\begin{equation}\label{sym3}
\mbox{ln}[R(I+4,I,A)]=[(\mu_n-\mu_p)-8(I+2)a_{sym}/A+2a_c(2Z-3)/A^{1/3}]/T+\Delta(I+4,I,A).\\
\end{equation}

Taking isobars with \textit{"I-2"} as the reference nuclei, the difference between $2\mbox{ln}[R(I+2,I,A)]$ and $\mbox{ln}[R(I+4,I,A)]$ can be written as
\begin{equation}\label{sym4}
(8a_{sym}/A+2a_c/A^{1/3})/T=\mbox{ln}[R(I+2,I,A)]-\mbox{ln}[R(I+4,I,A)]-\Delta(I+2,I,A)+\Delta(I+4,I,A).\\
\end{equation}
It can be found that
\begin{eqnarray}
\mbox{ln}[R(I+2,I,A)]-\mbox{ln}[R(I+4,I,A)]+\mbox{ln}[R(I+4,I+2,A)]\nonumber\\
-\Delta(I+4,I+2,A)-\Delta(I+2,I,A)+\Delta(I+4,I,A)=0.
\end{eqnarray}

Similarly, taking isobars with \textit{"I-2"} as the reference
nuclei, the difference between
$2\mbox{ln}[R(I,I-2,A)]-\mbox{ln}[R(I+2,I,A)]$ reads
\begin{equation}\label{eq.muac}
[(\mu_n-\mu_p)+2a_cZ/A^{1/3}]/T=2\mbox{ln}[R(I,I-2,A)]-\mbox{ln}[R(I+2,I,A)]-2\Delta(I,I-2,A)+\Delta(I,I+2,A).
\end{equation}

In Ref.~\cite{Huang10}, $[(\mu_n-\mu_p)]/T$ for different reaction systems is expressed as $[(\mu_n-\mu_p)/T]_i=[(\mu_n-\mu_p)/T]_0+\Delta\mu(Z/A)/T$, in which $(Z/A)=(Z_p+Z_t)/(A_p+A_t)$, and $p$ and $t$ represent the projectile and the target nuclei. Taking one reaction system as the reference and fitting the different reaction systems, $\Delta\mu(Z/A)/T$ for each reaction system can be fixed. Then $a_{sym}/T$ and $a_c/T$ are extracted from
the values of $(\mu_n-\mu_p)/T$ for each reaction system. Here, $a_{sym}/T$, $a_c/T$ and $(\mu_n-\mu_p)/T$ can be extracted
using Eq.~(\ref{sym2}), (\ref{sym4}), and (\ref{eq.muac})
only for one projectile fragmentation reaction.

\begin{figure*}[htbp]
%\begin{minipage}[t]{8.6cm}
\centering\includegraphics%[width=\columnwidth]
[width=8.6cm]{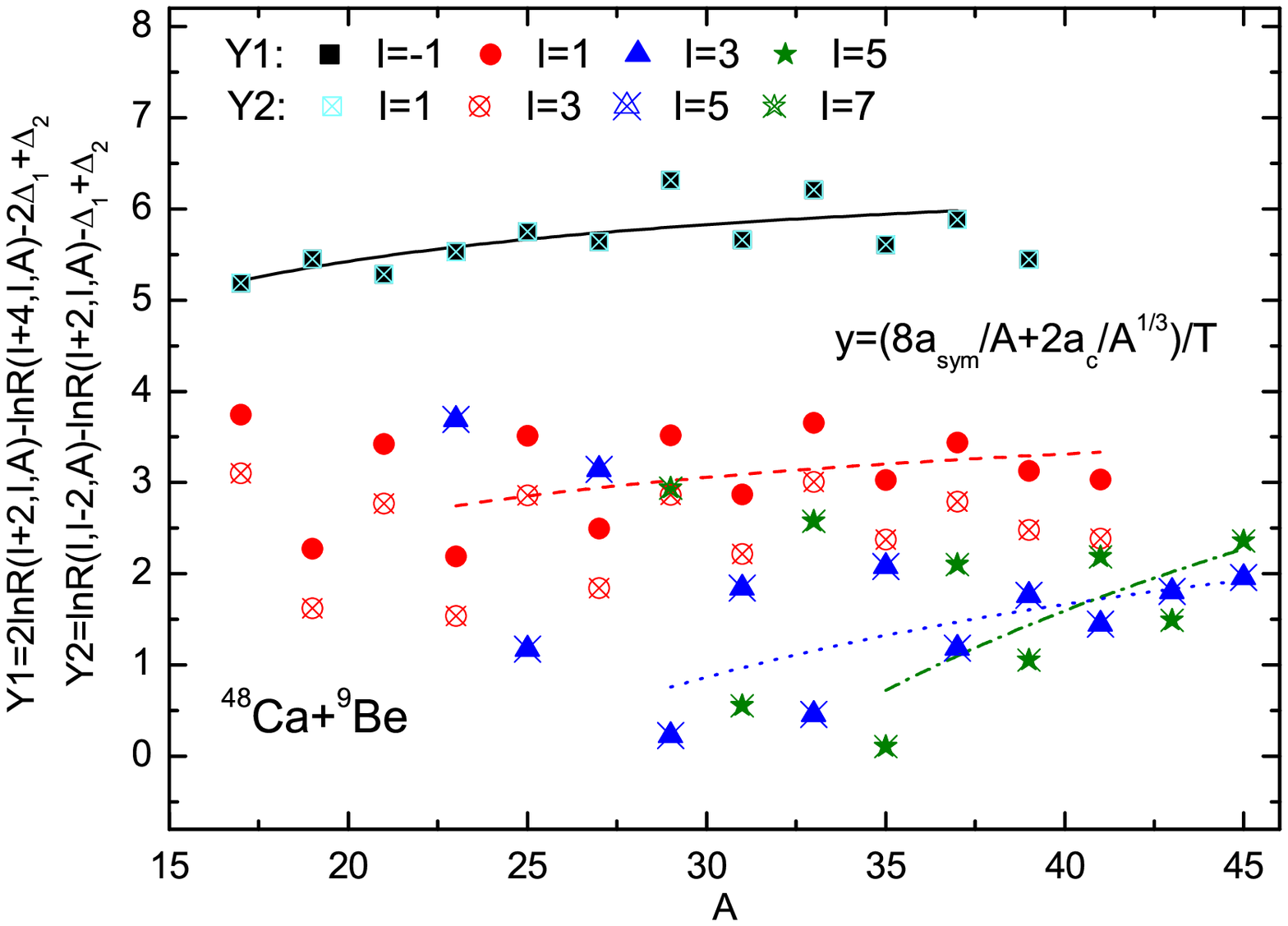}\caption{\label{Ca48asymac} (Color
online) Correlations between $(8a_{sym}/A+2a_c/A^{1/3})/T$ for
different isobars with \textit{I} and $A$ of fragments in the 140
$A$ MeV $^{48}$Ca + $^9$Be reaction. The lines are the fitting
results using Eq.~(\ref{sym2}).}
%\end{minipage}\hskip .5cm%
%\begin{minipage}[b]{8.6cm}
\centering\includegraphics%[width=\columnwidth]
[width=8.6cm]{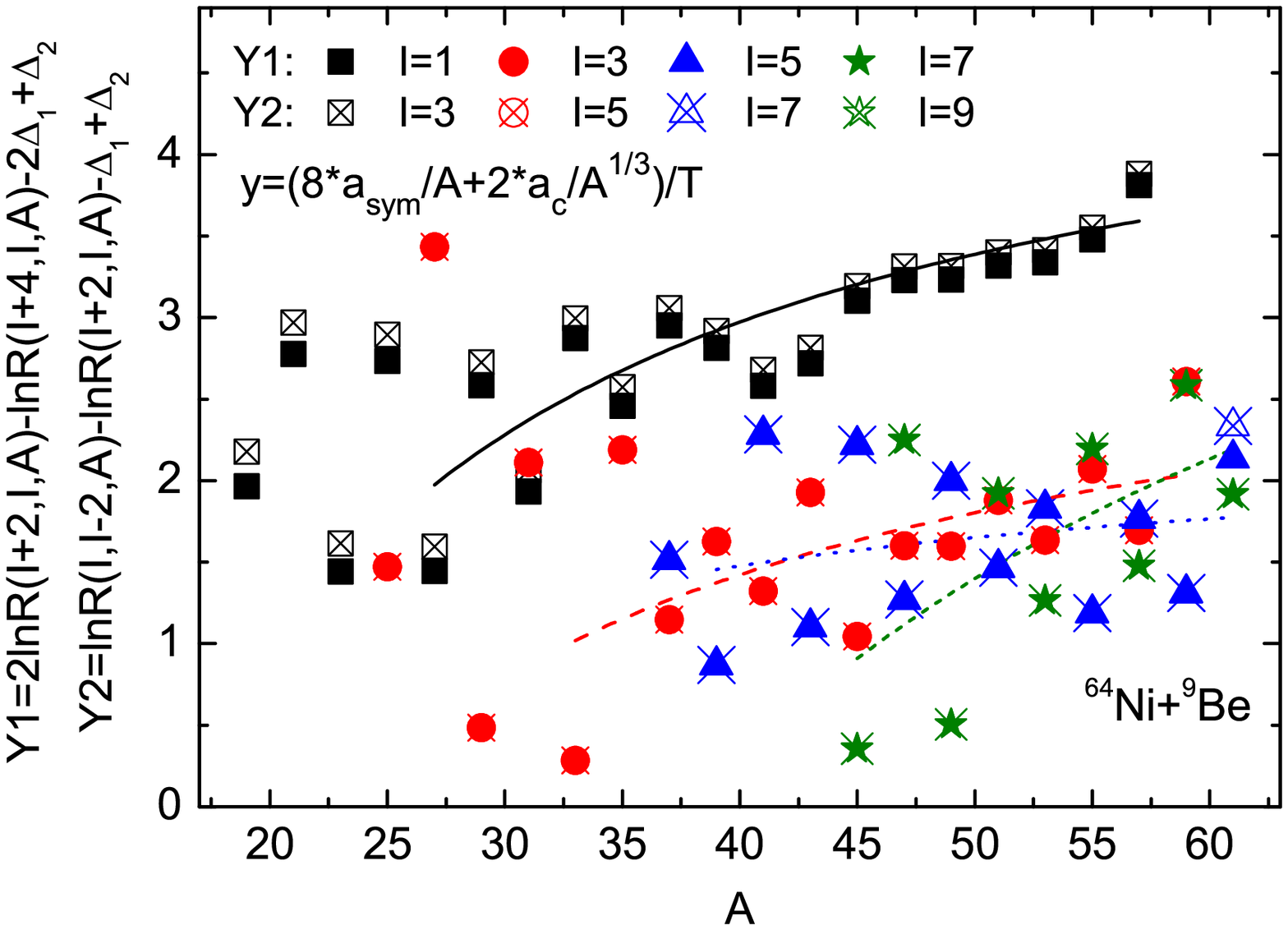}\caption{\label{Ni64asymac} (Color online) Correlations between $(8a_{sym}/A+2a_c/A^{1/3})/T$ for different isobars with \textit{I} and
A of fragments in the 140 $A$ MeV $^{64}$Ni + $^9$Be reaction. The lines are the fitting results using Eq.~(\ref{sym2}).}
%\end{minipage}
\end{figure*}

Using Eq.~(\ref{Rmirror1}), (\ref{sym2}), (\ref{sym4}),
and (\ref{eq.muac}), we analyze the yield ratios of isobars
produced in the 140$A$ MeV $^{48}$Ca + $^9$Be and $^{64}$Ni +
$^9$Be reactions~\cite{Mocko1,Mocko2}. In Fig.~\ref{Ca48asymac} and
~\ref{Ni64asymac}, the correlations between
$\mbox{ln}[R(I,I-2,A)]-\mbox{ln}[R(I+2,I,A)]$ and the mass number
($A$) of fragments [Eq.~(\ref{sym2})], and the correlations
between $\mbox{ln}[R(I+2,I,A)]-\mbox{ln}[R(I+4,I,A)]$ and $A$ of
fragments [Eq.~(\ref{sym4})] are plotted. These correlations are
fitted using a function $y =(8a_{sym}/A + 2a_c/A^{1/3})/T$, in
which $A$ is the argument and $a_{sym}/T$ and $a_c/T$ are parameters.
The values for $\mbox{ln}[R(1,-1,A)]-\mbox{ln}[R(3,1,A)]$ shows a large
difference to the values for
$\mbox{ln}[R(I,I-2,A)]-\mbox{ln}[R(I+2,I,A)]$. For isobars with
$I\geq3$, the values for
$\mbox{ln}[R(I,I-2,A)]-\mbox{ln}[R(I+2,I,A)]$ overlap.

\begin{figure*}[htbp]
%\begin{minipage}[t]{8.6cm}
\centering\includegraphics%[width=\columnwidth]
[width=8.6cm]{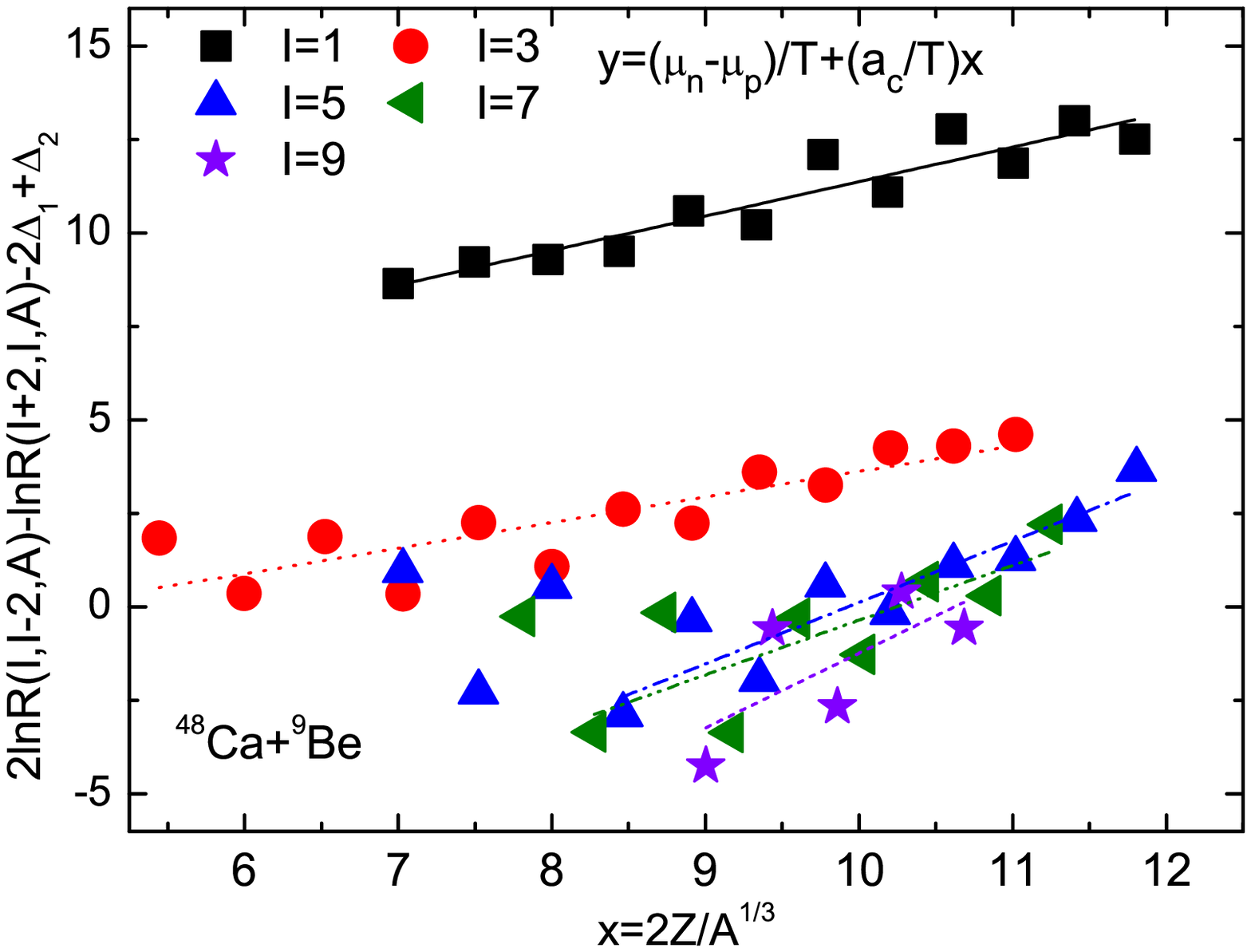}\caption{\label{Ca48muac} (Color online) Correlations between $2\mbox{ln}[R(I,I-2,A)]-\mbox{ln}[R(I+2,I,A)]$ and $2Z/A^{1/3}$ of
fragments produced in the 140 $A$ MeV $^{48}$Ca + $^9$Be reaction. The lines are the fitting results using Eq.~(\ref{eq.muac}).}
%\end{minipage}\hskip .5cm%
%\begin{minipage}[b]{8.6cm}
\centering\includegraphics%[width=\columnwidth]
[width=8.6cm]{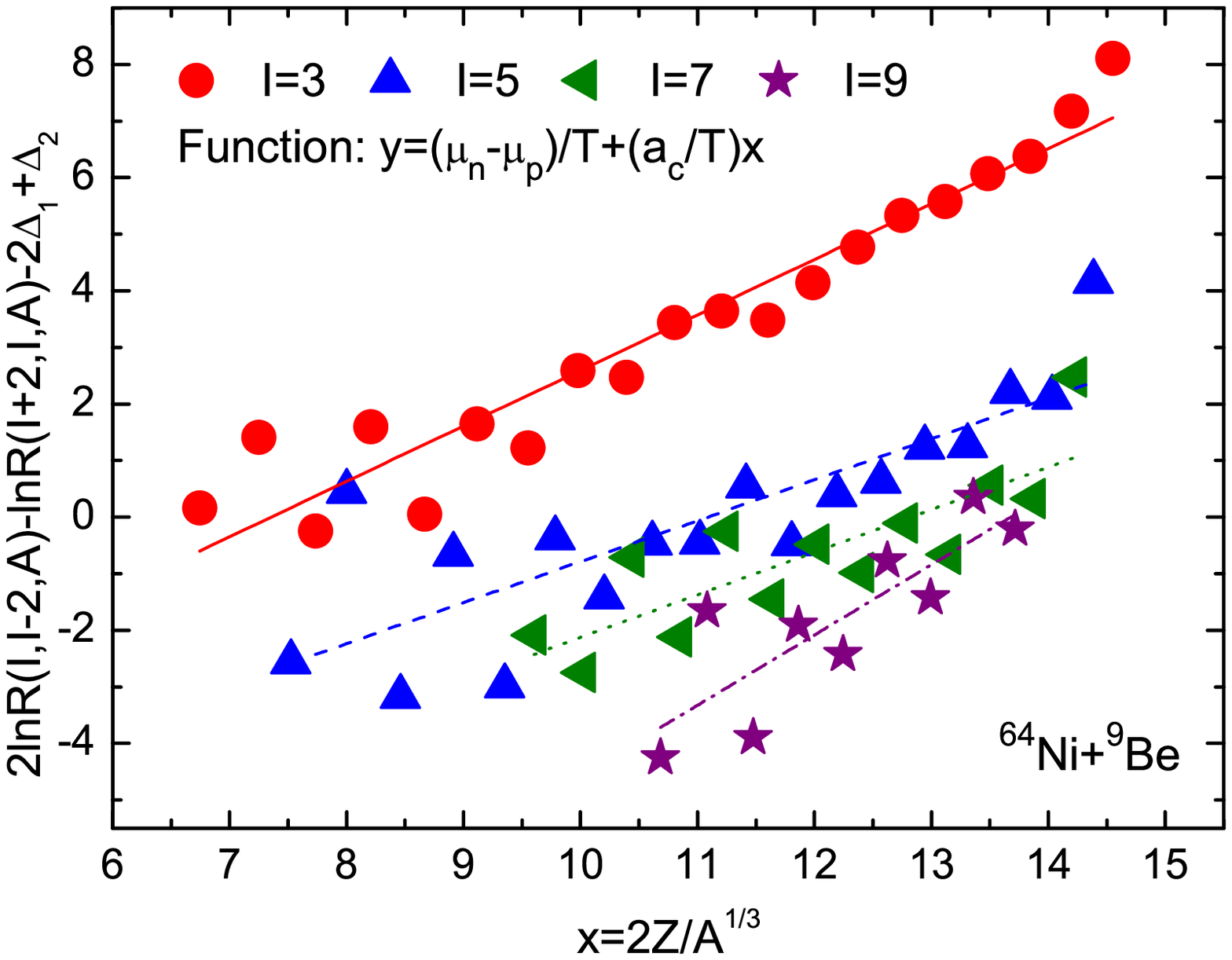}\caption{\label{Ni64muac} (Color online) Correlations between $2\mbox{ln}[R(I,I-2,A)]-\mbox{ln}[R(I+2,I,A)]$ and $2Z/A^{1/3}$ of
fragments produced in the 140 $A$ MeV $^{64}$Ni + $^9$Be reaction. The lines are the fitting results using Eq.~(\ref{eq.muac}).}
%\end{minipage}
\end{figure*}

In Fig.~\ref{Ca48muac} and \ref{Ni64muac},  the correlations
between $2\mbox{ln}[R(I,I-2,A)]-\mbox{ln}[R(I+2,I,A)]$  and
$2Z/A^{1/3}$ of isobars (using Eq.~\ref{eq.muac}) are depicted.
The correlations are fitted using a function
$y=[(\mu_n-\mu_p)+2a_cZ/A^{1/3}]/T$. The values for
$[(\mu_n-\mu_p)+2a_cZ/A^{1/3}]/T$ for isobars with \textit{"I=-1"}
and \textit{"I = 1"} shows large differences, but these decrease and
there is an overlap as \textit{I} increases.

\begin{figure*}[htbp]
%\begin{minipage}[t]{8.6cm}
\centering\includegraphics%[width=\columnwidth]
[width=8.6cm]{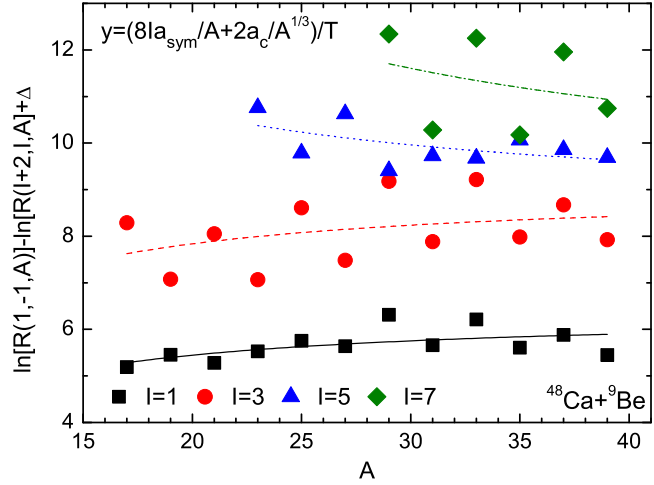}\caption{\label{Ca48mirror}
(Color online)  Correlations between
$\mbox{ln}[R(1,-1,A)]-\mbox{ln}[R(I+2,I,A)]$ for isobars  with
different \textit{I} and $A$ of fragments in the 140 $A$ MeV
$^{48}$Ca + $^9$Be reactions. The lines are the fitting results
using Eq.~(\ref{eq.asymac2}).}
\end{figure*}

In Fig.~\ref{Ca48mirror},  the correlations between
$\mbox{ln}[R(1,-1,A)]-\mbox{ln}[R(I+2,I,A)]$ for isobars with
\textit{I} and $2Z/A^{1/3}$ (using Eq.~(\ref{eq.asymac2})) for
fragments are displayed. The correlations are fitted using
Eq.~(\ref{eq.muac}). The values for
$[(\mu_n-\mu_p)+2a_cZ/A^{1/3}]/T$ for isobars with \textit{"I = -1"}
and \textit{"I = 1"} show large differences, but these decreases and there is an
overlap as \textit{I} increases.

Omitting the difference  between $(\mu_n-\mu_p)/T$ for nuclei with
different \textit{I}, and taking isobars with \textit{"I = -1"} as the
reference nuclei, the difference between
$\mbox{ln}[R(1,-1,A)]-\mbox{ln}[R(I+2,I,A)]$ can be written as
\begin{equation}\label{eq.asymac2}
(8Ia_{asym}/A+2a_c/A^{1/3})/T=\mbox{ln}[R(1,-1,A)]-\mbox{ln}[R(I+2,I,A)]+\Delta(I,I+2,A).
\end{equation}
In Fig.~\ref{Ca48mirror},  the correlations between
$\mbox{ln}[R(1,-1,A)]-\mbox{ln}[R(I+2,I,A)]$ for isobars with
different \textit{I} and $A$ for fragments are plotted. The
correlations are fitted using Eq.~(\ref{eq.asymac2}). The values
for $\mbox{ln}[R(1,-1,A)]-\mbox{ln}[R(I+2,I,A)]$ for different
isobars increase as the \textit{I} of isobars increases.

\begin{figure*}[htbp]
%\begin{minipage}[t]{8.6cm}
\centering\includegraphics%[width=\columnwidth]
[width=8.6cm]{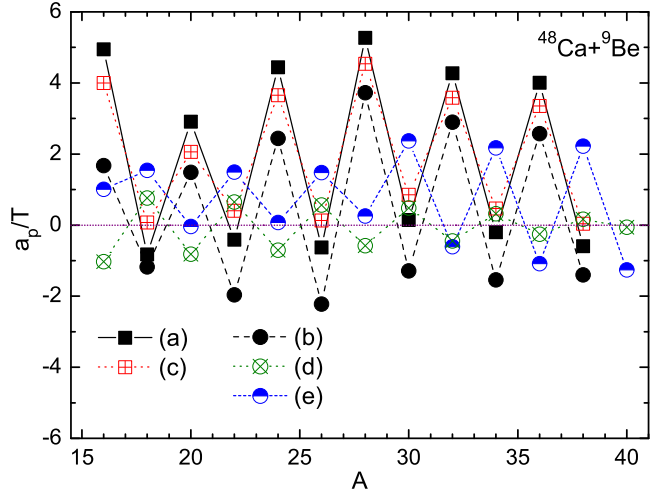}\caption{\label{Ca48paring} (Color
online) Correlations between $a_p/T$ and $A$ for fragments produced
in the 140 $A$ MeV $^{48}$Ca + $^9$Be reaction. (a), (b), (c), (d), and (e) are results obtained using Eqs.~(\ref{ap02_1}), (\ref{ap24_1}), (\ref{ap02_2}), (\ref{ap24_2}), and (\ref{ap24_3}), respectively.}
\end{figure*}

Following the same methods  in Ref.~\cite{Huang10}, the pairing
term $a_p/T$ for isobars with \textit{I=0} and \textit{I=2} are
rewritten here:
\begin{equation}\label{ap02_1}
a_p/T\sim(sgn)\frac{1}{2}A^{1/2}\{\mbox{ln}[R(2,0,A)]-\frac{1}{2}\{\mbox{ln}[R(1,-1,A)]+\mbox{ln}[R(3,1,A)]-\Delta(3,1,A)\}-\Delta(2,0,A)\},
\end{equation}
and for isobars with $I=2$ and $I=4$,
\begin{equation}\label{ap24_1}
a_p/T\sim(sgn)\frac{1}{2}A^{1/2}\{\mbox{ln}[R(4,2,A)]-\frac{1}{2}\{\mbox{ln}[R(1,-1,A)]-3\mbox{ln}[R(3,1,A)]+3\Delta(3,1,A)\}-\Delta(4,2,A)\}.
\end{equation}
Similarly, one can have
\begin{equation}\label{ap02_2}
a_p/T\sim(sgn)\frac{1}{2}A^{1/2}\{\mbox{ln}[R(2,0,A)]-\frac{1}{2}\{\mbox{ln}[R(1,-1,A)]-3\mbox{ln}[R(3,1,A)]+\Delta(3,1,A)\}-\Delta(2,0,A)\}
\end{equation}
and for isobars with $I=2$ and $I=4$,
\begin{eqnarray}\label{ap24_2}
a_p/T\sim(sgn)\frac{1}{2}A^{1/2}\{\mbox{ln}[R(4,2,A)]-\frac{1}{2}\{\mbox{ln}[R(3,1,A)-3\mbox{ln}[R(5,3,A)]]\nonumber\\
-\Delta(3,1,A)+3\Delta(5,3,A)\}-\Delta(4,2,A)\},
\end{eqnarray}
and
\begin{eqnarray}\label{ap24_3}
a_p/T\sim(sgn)\frac{1}{2}A^{1/2}\{\mbox{ln}[R(4,2,A)]-\frac{1}{2}\{\mbox{ln}[R(3,1,A)]+\mbox{ln}[R(5,3,A)]\nonumber\\
+\Delta(3,1,A)+\Delta(5,3,A)\}-\Delta(4,2,A)\}.
\end{eqnarray}
For an (odd,odd) nucleus $sgn=1$ and for an (even,even) nucleus
$sgn=-1$. The approximations assumed  in Eq.(\ref{ap02_1}) and
 (\ref{ap24_1}) are $(\mu_n-\mu_p)/T$, $a_{sym}/T$ and $a_c/T$
in ln$[R(3,1,A)]$ and ln$[R(4,2,A)]$ and are the same as those in
ln$[R(1,-1,A)]$. Similar approximations are made in
Eq.~(\ref{ap02_2}), (\ref{ap24_2}) and  (\ref{ap24_3}). In
Fig.~\ref{Ca48paring}, correlations between $a_p/T$ and $A$ of
fragments produced in the 140 $A$ MeV $^{48}$Ca + $^9$Be reactions
are plotted. $a_p/T$ of isobars with $I = 0$ and $I = 2$ are
extracted using Eq.~(\ref{ap02_1})-(\ref{ap24_3}),
respectively. In Eq.(\ref{ap02_1}), the chemical term, symmetry
term and Coulomb term of $R(2,0,A)$ are assumed to be equal to
those of $R(1,-1,A)$. In Eq.~(\ref{ap02_2}), $(\mu_n-\mu_p)/T$,
$a_{sym}/T$ and $a_c/T$ of R(2,0,A) are assumed to be equal to
those of $R(3,1,A)$. For (even,even) isobars with \textit{I=0} and
\textit{I=2}, the extracted $a_p/T$ using Eq.~(\ref{ap02_1}) is
bigger than that using Eq.~(\ref{ap02_2}), while for (odd,odd)
isobars the extracted $a_p/T$ using Eq.~(\ref{ap02_1}) are smaller
than those using Eq.~(\ref{ap02_2}).

Due to lack of data for cross sections of mirror
nuclei in the $^{64}$Ni projectile fragmentation, figures like
Figs.~\ref{Ca48mirror} and \ref{Ca48paring} are not plotted
for $^{64}$Ni. From Figs.~\ref{Ca48asymac} to
 \ref{Ca48paring}, it can be seen that the correlations
between isobaric yield ratios and $(\mu_n-\mu_p)/T, a_{sym}/T$,
$a_c/T$, and $a_p/T$ can fit the measured data well. But for
isobars with big $I$, there are not enough data to form
chains and it is impossible to extract the values for
$a_{sym}/T$ and $a_c/T$ for these very neutron-rich isobars.

The method discussed above has the great advantage that the analysis can be performed in a single reaction and there is no need to calibrate $\Delta\mu(Z/A)$ as in Ref. \cite{Huang10}. The extracted values of $(\mu_n-\mu_p)/T, a_{sym}/T$, $a_c/T$, and $a_p/T$ are at a specific temperature associated with the incident energy. $(\mu_n-\mu_p)/T, a_{sym}/T$, $a_c/T$, and $a_p/T$ are all temperature dependent. To study the dependence of $(\mu_n-\mu_p)/T, a_{sym}/T$, $a_c/T$, and $a_p/T$ on temperature, projectile fragmentation at different energies should be investigated.

\section{Summary}
In summary, the coefficients of the Coulomb  energy $a_{c}/T$, symmetry
energy $a_{sym}/T$, pairing energy $a_p/T$, and $(\mu_n-\mu_p)/T$
have been studied by analyzing the yield ratios ($R$) of
isobars in projectile fragmentation in the framework of the
modified Fisher model. Very simple correlations between
$(\mu_n-\mu_p)/T$, $a_c/T$, $a_{sym}/T$, $a_p/T$, and $R$ are
obtained. It is found that these correlations can fit the experimental results
well and can be used to extract the symmetry energy of the neutron-rich nuclei.

\begin{acknowledgments}
This work is partially supported by the National Natural Science
Foundation of China under Contract No. 10905017, and No. 11035009, the
Program for Innovative Research Team (in Science and Technology)
under Grant No. 2010IRTSTHN002 in the University of Henan Province,
China, and the Shanghai Development Foundation for Science and
Technology under Contract No. 09JC1416800, and the Knowledge
Innovation Project of the Chinese Academy of Sciences under Grant
No. KJCX2-EW-N01.
\end{acknowledgments}

\end{document}